\documentclass[12pt,preprint]{aastex}
\bibpunct{(}{)}{;}{a}{}{,}

\shorttitle{A 42.-43.6~GHz spectral survey of Orion BN/KL}
\shortauthors{Goddi et al.}

\newcommand{\tr}{$^{30}$SiO}
\newcommand{\tn}{$^{29}$SiO}
\newcommand{\te}{$^{28}$SiO}

\newcommand{\HII}{H{\sc ii}}

\newcommand{\kms}{km~s$^{-1}\,$}

\newcommand{\pas}{.\hskip-2pt$^{\prime\prime}$}

\begin{document}

\title{A 42.3-43.6~GHz spectral survey of Orion BN/KL:\\
First detection of the $v=0 \ J=1-0$ line from the isotopologues \tn~and \tr.}

\author{C. Goddi, L. J. Greenhill, E. M. L. Humphreys, L. D. Matthews}
\affil{Harvard-Smithsonian Center for Astrophysics,
    60 Garden Street, Cambridge, MA 02138}
\and

\author{Jonathan C. Tan}
\affil{Department of Astronomy, University of Florida, Gainesville, FL 32611}

\and

\author{C. J. Chandler}
\affil{National Radio Astronomy Observatory, P.O. Box O, Socorro, NM 87801}

\begin{abstract}
We have surveyed molecular line emission from Orion BN/KL from 42.3 to 43.6 GHz with the Green Bank Telescope. Sixty-seven lines were identified and ascribed to 13 different molecular species. The spectrum at 7~mm is dominated by SiO, SO$_2$, CH$_3$OCH$_3$, and C$_2$H$_5$CN.  
Five transitions have been detected from the SiO isotopologues \te, \tn, and \tr. 
 We report here for the first time the spectra of the \tn~and \tr~$v=0 \ J=1-0$ emission in Orion BN/KL, and we show that they have  double-peaked profiles with  velocity extents similar to the main isotopologue. 
The main motivation for the survey was the search of high-velocity (100-1000~\kms) outflows in the BN/KL region as traced by SiO Doppler components. Some of the unidentified lines in principle could be  high-velocity SiO features, but without imaging data their location cannot be established.
 Wings of emission are present in the $v=0$ \te, \tn~and \tr~profiles, and we suggest that the $v=0$ emission from the three isotopologues might trace a moderately high-velocity ($\sim 30-50$~\kms) component   of the flows around the high-mass protostar Source\,I in the Orion BN/KL region.
 We also confirm the 7~mm detection of a complex oxygen-bearing species, acetone (CH$_3$COCH$_3$), which has
 been recently observed  towards the hot core at 3~mm, and we have found further indications of the presence of long cyanopolyynes (HC$_5$N and HC$_7$N) in the quiescent cold gas of the extended ridge.
\end{abstract}


\section{Introduction}
\label{int}

The luminous ($L\approx 10^{5}~L_{\odot}$)  Orion BN/KL infrared nebula is the nearest ($D \sim 414$~pc: \citealt{Men07}) and probably most studied high-mass star forming region (HMSFR) in the Galaxy.
More than 20 spectral line surveys have been carried out in the mm and submm bands over the last 25 years (72-91 GHz: \citealt{Joh84}; 70-115 GHz: \citealt{Tur89}; 138-151~GHz: \citealt{Lee01}; 150-160~GHz: \citealt{Ziu93}; 160-165~GHz: \citealt{Lee02}; 215-247~GHz: 
\citealt{Sut85}; 218-243~GHz: \citealt{Bla96}; 247-263~GHz: \citealt{Bla86}; 257-273~GHz: \citealt{Gre91}; 330-360~GHz: \citealt{Jew89}; 325-360~GHz: \citealt{Sch97}; 342-359~GHz: \citealt{Whi86}; 334-343~GHz: \citealt{Sut95}; 455-507~GHz: \citealt{Whi03}; 486-492 and 541-577~GHz: \citealt{Olo07}; 607-725~GHz: \citealt{Sch01}; 795-903~GHz: \citealt{Com05}).  
Imaging line surveys of Orion-BN/KL have also been conducted using the Submillimeter Array interferometer (SMA) in the submm ranges 337.2-339.2~GHz and 347.2-349.2~GHz \citep{Beu05}, and 679.78-681.75~GHz and 689.78-691.75~GHz \citep{Beu06}. 
 
 We report here the first high spectral resolution survey at 7~mm between frequencies of 42.3 and 43.6~GHz. The main target of our survey was SiO which is a well-known tracer of shocks  in HMSFRs, and so it is particularly suitable for tracing  high-velocity outflowing molecular gas. 
 In this context, SiO observations provide observational constraints on the acceleration mechanisms of high-mass protostellar outflows, and thus also test different theories of massive star formation. For example, if massive stars form from massive cores in a qualitatively similar but scaled-up fashion to low-mass stars \citep{Mck03} then similar disk and outflow properties are expected, such as magneto-centrifugally launched X-winds \citep{Shu00} and/or disk-winds \citep{Kon00} that are aligned and collimated orthogonally to the plane of the accretion disk. Alternatively, if massive stars form via more chaotic accretion processes, such as competitive accretion \citep{Bon06} or stellar mergers \citep{Bon98}, then more disordered disks and outflows might be expected.
A generic prediction of X-wind and disk-wind models is that the outflow is initially accelerated to about the local escape speed of the launching region. Relatively large ($\sim$10-30\%) fractions of the accretion flow are predicted to be launched from the inner disk, within a few stellar radii, where the escape speeds can approach $\sim$1000~\kms (the escape speed from the surface of a 20~M$_{\odot}$ protostar that has contracted to the zero age main sequence, for example). However, such high-speed winds have never been observed close to high mass protostars.
 
Orion BN/KL is a good target for observations at 7~mm. First, strong SiO maser emission from three vibrational states are excited at 7~mm by a high-mass protostar in the region, Source~I \citep{Gre04}, and their exceptionally high brightness temperature might facilitate detection of (weak) high-velocity components in the protostellar wind  compared to thermal emission \citep{Gen81}. Second,  though Orion is a rich source of line emission at millimeter wavelengths \citep{Sut85,Sch97,Whi03}, one may reasonably anticipate a lower density of lines in the relatively long wavelength 7~mm band. Third, Source~I is the only known origin of SiO emission in Orion BN/KL \citep{Beu05}, diminishing the risk of confusion.

On the theoretical side, \citet{Tan03} presented models for magneto-centrifugally launched outflows from a massive protostar, such as the source that may be powering the Orion-BN/KL region, Source~I. 
\citet{Tan04} pointed out that close passage of the BN object, which is a runaway B star, may have tidally perturbed the accretion disk within the last 1000~yr, perhaps leading to enhanced accretion and outflow activity that may help to explain the ``explosive" appearance of the outflow on larger scales \citep{All93}. Alternatively, \citet{Bal05} have argued that the explosive, poorly-collimated nature of the outflow may have been caused by a protostellar merger. Observations of  fast (and inner) portions of the outflow, and their relation to any protostellar accretion disk, are essential to distinguish these various scenarios.

Thus, in this paper we searched for high velocity outflows  by using single-dish observations of SiO transitions. Although the beam ($16''$) of our observations is not sufficient to identify the driving source of the outflow in the region, previous high-angular resolution observations showed unequivocally that the SiO maser emission probes circumstellar gas at distances 10-1000~AU from Source~I (e.g., \citealt{Gre04}). While we found evidence for flows up to $\sim 50$~\kms, we did not find any conclusive indication of higher-speed winds, although several unidentified emission lines are candidates. As an additional byproduct of this search,
 we have detected for the first time  in Orion BN/KL the $v=0 \ J=1-0$ transition of two SiO isotopologues, \tn~and \tr, and we have identified a number of other chemical species in the region.

In this paper, we present the spectral profiles for the \te~$v=0,\ v=1,\ v=2 \ J=1-0$ and \tn~and \tr~$v=0 \ J= 1-0$ lines as well as all the molecular identifications in the surveyed range of frequencies. 
This work is structured as follows: \S 2 describes the observations and data reduction, the results of the line identification are illustrated in \S 3, and  an analysis on individual species is presented in \S 4. A summary is presented in \S 5.

\section{Observations and data reduction}
\label{obs}
The survey was conducted conducted with the National Radio Astronomy Observatory\footnote{The National Radio Astronomy Observatory  is a facility of the National Science Foundation operated under cooperative agreement by Associated Universities, Inc.} (NRAO) 100~m Robert Byrd Green Bank Telescope (GBT) on 2007 November 10 at 7~mm for a total of 4~h. A pointing position $\alpha_{2000} = 05^h 35^m 14$\pas5,   $\delta_{2000} = -05^{\circ} 22' 31''$, and an LSR source velocity of +5~\kms was used for Orion BN/KL. The telescope had a FWHM beamwidth of $16\arcsec$ at 7~mm.

The GBT spectrometer was configured to provide two  200~MHz IFs in two polarizations, with a frequency resolution of 24.4~kHz  (corresponding to a velocity resolution of $\sim 0.17$~\kms at 43~GHz). The two bands overlapped 40~MHz, resulting in an instantaneous band of 360~MHz.
 Four scheduling blocks of 1~h were observed at four different frequencies, that placed 
the following transitions of SiO near the band centers: \te~$J=1-0\ v=0$ (43423.76\,MHz), $v=1$ (43122.03\,MHz), $v=2$ (42820.48\,MHz), $v=3$ (42519.379\,MHz). 
The frequency setup resulted in a continuous coverage of 42.3-43.6~GHz (corresponding to a bandwidth of $\sim 9070$~\kms at 43~GHz).

Data were taken in a dual beam mode by nodding the antenna, with signal (on-source) and reference (off-source) spectra observed simultaneously in two beams, the target always being present in one. 
Antenna pointing and focus corrections were obtained every 45 minutes  by observing a strong nearby continuum source.  
Data were reduced using scripts written in the Interactive Data Language within the GBTIDL software. We removed 5th-order polynomial baselines  from the total-power spectra, which were subsequently boxcar smoothed to an effective resolution of 97.7~kHz (or 0.68~\kms).
Spectra were calibrated to units of flux density (Jy) using a value of zenith opacity of 0.07, estimated from measurements of T$_{\rm sys}$ at  different elevation angles ($10^{\circ}-70^{\circ}$),
and an aperture efficiency of $\eta_a=0.45$, based on the gain curve measured in 2005 over an elevation range $10^{\circ}-80^{\circ}$ \citep{Nik07}.
The resulting 1~$\sigma$ noise level attained in an integration of 30 minutes was $\sim 5.0$~mJy ($\sim 6.0$~mK in units of  corrected antenna temperature, $T_a^*$) in a 97.7~kHz channel. 
The units of peak intensity will be given throughout the paper either in Jy or K, where $S({\rm Jy}) = 0.78 T_a^*({\rm K})=0.46 T_{\rm MB}(K)$ is appropriate for the GBT at 43~GHz.

Because the \te~lines are quite strong (30-1000~Jy), we inspected the spectra for artifacts that might occur due to inefficiencies in the signal path (e.g., aliasing). We identified ``ghost" lines offset $\pm 100$~MHz from the $v=0$, 1, and 2 lines, attenuated by a factor of $\sim 400$ with respect to the line peaks. These artifacts arise from 100~MHz sidebands in the output of the Voltage Controlled Crystal Oscillator that is used to mix signals to a standard IF band at the GBT, and these therefore affect all observing bands \citep{Min07a,Min07b}. Other than the \te~lines, no other artifacts were detected; the ghosts associated with a typical 0.5~Jy line would be a factor of four below the observed noise level. 
For the purpose of this analysis we excised data in the contaminated frequency ranges. In total 30.9~MHz were lost.

%
\clearpage
\begin{figure*}
\centering
\includegraphics*[width=15cm]{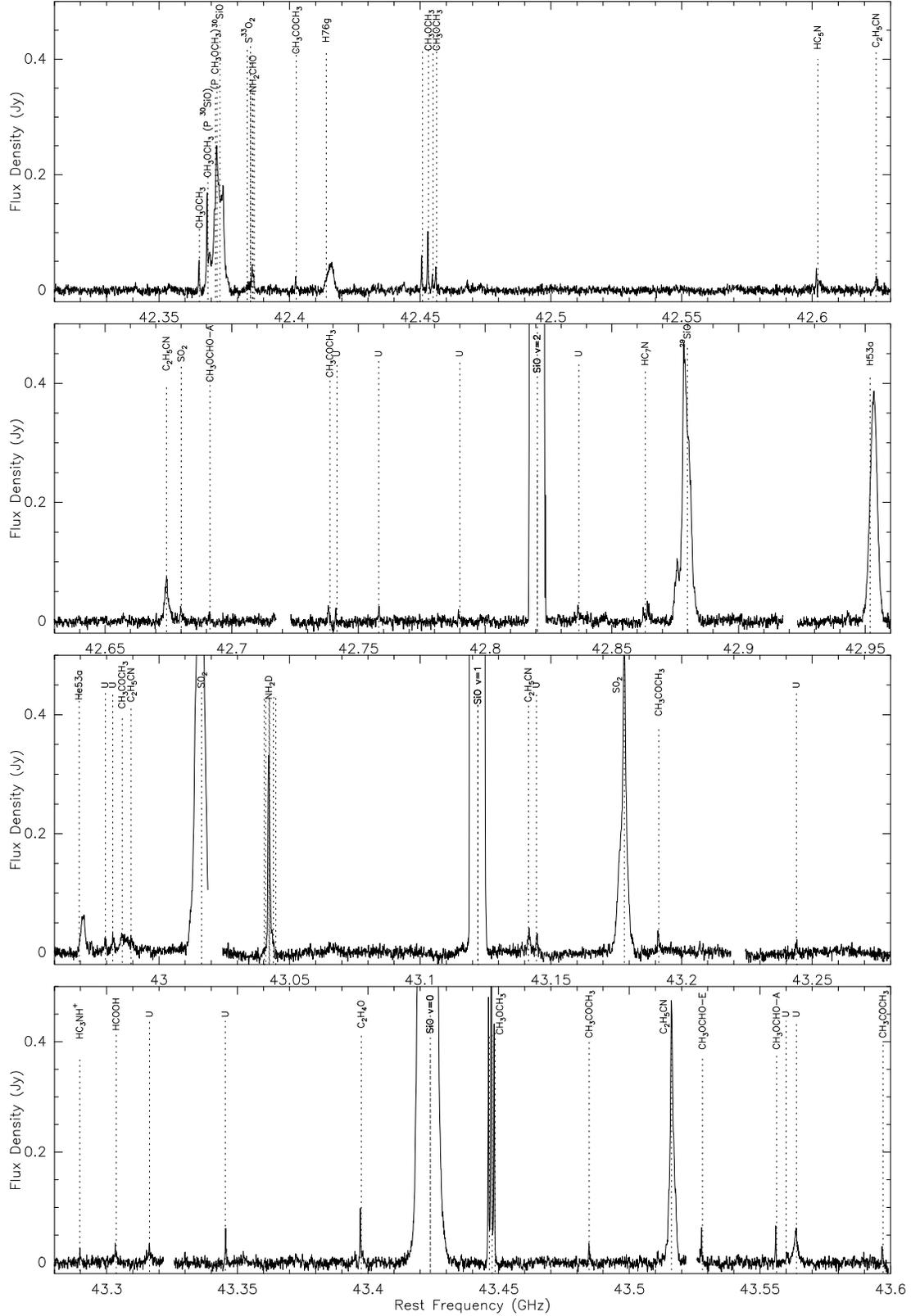}
\caption{Line survey from 42.1 to 43.6~GHz. All observed lines are marked. A systemic velocity $v_{\rm LSR} = 5$~\kms has been assumed and used to shift the frequency axis. The gaps in the spectra correspond to instrumental artifacts (see Sect.~\ref{obs}).}
\label{spec}
\end{figure*}
\clearpage
%
%

\section{Results of the line identification}

 Line identifications are based on the JPL \citep{Pic98}, CDMS \citep{Mul01}, and LOVAS \citep{Lov04}  catalogs.  
  Selection criteria include: frequency coincidence, line velocity, upper state energy, line strength,  and presence/absence of other lines from similar levels either in the surveyed regions or at higher frequencies.
  The calibrated spectra   are shown in Fig.~\ref{spec}. A $v_{\rm LSR} =5$~\kms has been used to derive the frequency axis and markers are placed at the laboratory rest frequencies of the transitions attributed to the observed line features.
Above 15~mJy (3$\sigma$), we found 55 resolvable features (12 unidentified) consisting of 67 lines, partially blended (Table~\ref{lines}).
We identified a total of 13 different species, 5 isotopologues, and 2 vibrationally excited states (from SiO) within the data set. Most of the detected lines can be attributed to known molecular species, previously detected in Orion BN/KL at higher frequencies: 
SiO, SO$_2$, NH$_2$D, HCOOH, NH$_2$CHO, C$_2$H$_5$CN, CH$_3$OCH$_3$, CH$_3$OCHO (Table~\ref{species}). 
Interestingly, in this survey we also confirm the detection of one complex oxygen-bearing species, CH$_3$COCH$_3$ (acetone), recently observed in Orion BN/KL at 3~mm (Sect.~\ref{ace}), and we found strong evidence for long carbon-chain (c-chain) molecules: HC$_5$N  and HC$_7$N (Sect.~\ref{cyano}), and possibly the cation HC$_3$NH$^+$ (Sect.~\ref{others}).

 Multiple velocity components are present in the Orion spectra \citep{Gen89}, which complicates the line identification: the hot core ($v_{\rm LSR}~$ $\sim 3-6$~\kms, $~\Delta v \sim 5-10$~\kms), the compact ridge $-$ warm and dense gas about 10$^{\prime\prime}$ south-west of the hot core ($v_{\rm LSR}~$ $\sim$ 8~\kms, $~\Delta v \sim$ 3~\kms), the extended ridge $-$ ambient gas in the Orion Molecular Cloud OMC-1 ($v_{\rm LSR}~\sim$ 9~\kms, $~\Delta v \sim 4$~\kms), and the plateau or outflow ($v_{\rm LSR}~\sim 6-10$~\kms, $~\Delta  v \sim 20$~\kms). 
 High-angular resolution ($<1''$) imaging \citep{Beu05,Beu06} unequivocally showed that different molecules are mostly concentrated in different regions and hence emit at different velocities: large oxygen-bearing species (such as CH$_3$OCH$_3$ and HCOOH) are found primarily toward the compact ridge and hence emit mostly at 8--10~\kms; large nitrogen-bearing species (e.g., C$_2$H$_3$CN and C$_2$H$_5$CN) and SiO are located toward the hot core and Source~I (separated by $\sim 1''$), respectively, and emit at $\sim 5$~\kms; finally, sulfur-bearing species (e.g., SO$_2$) show emission towards both the compact ridge and the hot core.

For all observed molecules, apparent shifts between the observed lines and the rest frequencies are evident (Fig.~\ref{spec}), corresponding to different origins among the several components of the star forming region. For example, oxygen-bearing species show an average shift from the laboratory frequencies around 0.5~MHz, which corresponds to $\Delta v \sim +4$~\kms from the assumed systemic velocity of 5~\kms, in agreement with an origin in the compact ridge. 

 When several lines of the same species (not affected by blending) were detected,  
 we performed a rotation diagram analysis to estimate column density, $N_{col}$, and rotational temperature, $T_{rot}$. With the assumption of LTE, the source-averaged column density from an optically thin transition is given by \citep{Sny05}:
\begin{equation}
\label{ntot}
 \frac{N_{col}}{Q} e^{-E_{u}/T_{rot}} = \frac{1.67 W}{B  \eta_{MB} S\mu^2\nu} \times 10^{14} \rm cm^{-2}  
\end{equation}
 where $Q$ is the partition function, $E_u$ is the upper state energy of the transition (K), $W$ is the integrated line intensity\footnote{The peak intensity is given by the corrected antenna temperature, $T_a^*$.} (K~\kms),  $B$  is the beam filling factor (see Eq.~2 of \citealt{Sny05}), $\eta_{MB} $ is the main beam efficiency ($\eta_{MB}=1.32 \times \eta_a$ for the GBT), $S\mu^2$ is the product of the line strength and the square of the molecular dipole moment (D$^2$), and $\nu$ is the transition frequency (GHz).
We obtain the rotation temperature and the column density from a least square fit of upper state energy  to integrated line intensity for different transitions in log space (a rotation diagram).
Different source sizes are assumed to calculate the beam filling factor, depending on the molecule and  origin (i.e., hot core, compact or extended ridge). 
For molecules with only one or a few blended lines, we used  temperature  and column density estimations from previous works \citep{Tur91,Sch97,Whi03}.

%
\clearpage
\begin{deluxetable}{lllll}
\tabletypesize{\footnotesize}
\tablewidth{0pc}
\tablecaption{Observed lines.}
\tablehead{
\colhead{$\nu$ (MHz) } & \colhead{Species} &
\colhead{Transition} & \colhead{$E_{u}/k$(K)} & \colhead{Notes} 
}
\startdata
42365.574   &CH$_3$OCH$_3$ &   $13_{3,11}AA -  12_{4,8}AA $ & 96 &  \\    
42368.723   & CH$_3$OCH$_3$ & $13_{3,11}EE -  12_{4,8}EE $ & 96& O \\    
42371.579   & CH$_3$OCH$_3$ & $13_{3,11}EA -  12_{4,8}EA $ & 96& O  \\    
42372.164   &CH$_3$OCH$_3$ &  $13_{3,11}AE -  12_{4,8}AE $ & 96& O \\    
42373.340    & $^{30}$SiO & $1-0$ & 2 & \\
   42383.800    & $^{33}$SO$_2$    &   $21_{3,19} - 20_{4,16}$ & 234 &\\  
   42385.100    & $^{33}$SO$_2$    &  $21_{3,19} - 20_{4,16}$ & 234 &\\  
   42385.061 & NH$_2$CHO & $2_{0,2},1-      1_{0,1},1 $ & 3 &  \\   
   42385.655 & NH$_2$CHO & $2_{0,2},1-      1_{0,1},2 $  & 3 &   \\   
   42386.070 & NH$_2$CHO & $2_{0,2},2-      1_{0,1},1 $  & 3 &   \\     
   42386.547 & NH$_2$CHO & $2_{0,2},1-      1_{0,1},0 $  & 3 &   \\   
   42386.681 & NH$_2$CHO & $2_{0,2},2-      1_{0,1},2 $   & 3 &  \\  
   42402.583   &CH$_3$COCH$_3$ &  $11_{9,2} -  11_{8,3} $ & 58 & \\ 
     42414.0& H76$\gamma$ & &  \\
42450.801        &CH$_3$OCH$_3$ &  $16_{4,13}AA-15_{5,10}AA $ &147 & \\  
42453.197        &CH$_3$OCH$_3$ &  $16_{4,13}EE-15_{5,10}EE $ &147& \\  
42454.906        &CH$_3$OCH$_3$ &  $16_{4,13}EA-15_{5,10}EA $ &147&   \\       
42456.335 & CH$_3$OCH$_3$ &  $16_{4,13}AE-15_{5,10}AE $ &147 &   \\     
42602.153& HC$_{5}$N &$16-15$ & 17& P \\
42624.488 & C$_2$H$_5$CN&  $23_{3,21}-22_{4,18}$ & 129 & \\
42674.214& C$_2$H$_5$CN&  $11_{1,10}-11_{0,11}$ & 30 & \\
42680.090 & SO$_2$ & $41_{7,35}-40_{8,32}$ & 913 & \\
42691.505 & CH$_3$OCHO-A & $19_{11,8}-20_{10,11}$ & 192 &\\
42738.766    & CH$_3$COCH$_3$ &  $11_{8,4}-11_{7,5} $ & 55 & \\   
42741.533 & ... & ... &... &U\\ 
42758.016 & ... &... &... & U \\
42790.000 &...&...& ...& U \\
42820.480& SiO $v=2$ & $1-0$& 2 &\\
42836.975 & ... & ... &... & U\\
42863.209 & HC$_7$N& $38_{38}-37_{37}$ & 40 & P \\
42879.846 &$^{29}$SiO& $1-0$ & 2 &\\
42951.970 & H53 $\alpha$& & \\
42968.75 & He53 $\alpha$ &  &  \\
42979.467 &...&... & ...& U\\
42982.400 &...&... &... & U\\
42985.958 &CH$_3$COCH$_3$ &  $11_{8,4} -  11_{7,5} $ & 55  & \\ 
42989.389 & C$_2$H$_5$CN & $13_{11,2}-13_{11,3}$& 41 & \\
43016.280&SO$_2$& $19_{2,18}-18_{3,15}$& 181 & \\
43040.308 &NH$_2$D& $3_{1,3},1_2-3_{0,3},0_3$& 95 &\\  
43040.856 &NH$_2$D& $3_{1,3},1_4-3_{0,3},0_3$&95 &\\  
43042.160  &NH$_2$D& $3_{1,3},1_2-3_{0,3},0_2$&95 &\\
43042.228   &NH$_2$D& $3_{1,3},1_4-3_{0,3},0_4$&95 &\\  
43042.421   &NH$_2$D& $3_{1,3},1_3-3_{0,3},0_3$&95 &\\
43043.793   &NH$_2$D& $3_{1,3},1_3-3_{0,3},0_4$&95 &\\
43044.273   &NH$_2$D& $3_{1,3},1_3-3_{0,3},0_2$&95 &\\  
43122.080 &SiO $v=1$& $1-0$& 2 &\\
43141.412 & C$_2$H$_5$CN  & $7_{3,5}-8{2,6}$& 22 \\
43145.898 & ... & ...&... & U\\
43178.140&SO$_2$& $23_{2,22}-22_{3,19}$& 258 &\\
43191.308 & CH$_3$COCH$_3$ & $10_{5,5}-10_{4,6}$ & 43 & \\
43244.00 &...& ...& ...& U \\
43289.744 & HC$_3$NH$^+$ & $5-4$& 4& P\\
43303.710& HCOOH & $2_{1,2}-1_{1,1}$ & 6 &\\
43316.000 &...&... &... & U \\
43345.500 &...& ... &... & U \\
43397.471  & c-C$_2$H$_4$O & $8_{6,2}-8_{5,3}$& 78 & P\\
43423.760 & SiO $v=0$ & $1-0$ & 2 & \\
43446.471 &CH$_3$OCH$_3$ &  $6-{1,5}AE -6_{0,6}AE$ & 21 &\\  
43446.471  &CH$_3$OCH$_3$ &  $6-{1,5}EA -6_{0,6}EA$& 21 & \\  
43447.542  &CH$_3$OCH$_3$ &  $6-{1,5}EE -6_{0,6}EE$ & 21 &\\  
43448.612  &CH$_3$OCH$_3$ &  $6-{1,5}AA -6_{0,6}AA$ & 21 &\\ 
43484.704 &CH$_3$COCH$_3$ &  $13_{10,3}-13_{9,4} $ & 80 & \\   
43516.205 & C$_2$H$_5$CN  & $5_{1,5}-4{1,4}$& 7 & \\ 
43528.060 & CH$_3$OCHO-E &$7_{1,6}-7_{1,7}$ & 18 & \\
43556.436 & CH$_3$OCHO-A &$7_{1,6}-7_{1,7}$ & 18& \\
43560.000& ... & ... &... & U \\
43564.000&...&... & ...& U \\
43597.127 &CH$_3$COCH$_3$ &  $4_{0,4}-3_{1,3} $ & 6\\  
& & & & \\
\hline 
\enddata
\tablecomments{
\scriptsize 
U = Unidentified feature; P = Possible detection; O = line overlapped with the \tr~profile.\\
Hyperfine components are listed separately if spaced by $> 0.1$~MHz.}
\label{lines}
\end{deluxetable}
\clearpage
%

\begin{table}
\centering
\caption {Detected Molecular Species}
\begin{tabular}{ccc}
\hline\hline
Species & Isotopologues/Isotopomers & Lines\tablenotemark{a} \\
\hline
$^{28}$SiO & $^{29}$SiO,$^{30}$SiO& 5 \\
SO$_2$ & $^{33}$SO$_2$& 5 \\
CH$_3$OCH$_3$ & & 12 \\
CH$_3$COCH$_3$ && 6 \\
C$_2$H$_5$CN && 5 \\
&NH$_2$D & 7 \\
NH$_2$CHO && 6 \\
HCOOH &  & 1 \\
CH$_3$OCHO & & 3 \\
& c-C$_2$H$_4$O  & 1\\
HC$_5$N   && 1 \\ 
HC$_7$N && 1 \\
HC$_3$NH$^+$  && 1 \\
\hline
\end{tabular}
\tablenotetext{a}{The total number of lines for each species is based on the list in Table~\ref{lines}.}
\label{species}
\end{table}
\clearpage
%
\section{Analysis of individual species}
In this section, we discuss individual species, focusing on 
SiO, the main target of this survey, and a few other species of interest.

\subsection{SiO}

The $v=0, v=1, v=2\ J=1-0$ transitions of \te, as well as the $v=0\ J=1-0$ lines of \tn~and \tr~\ were detected in the survey (Fig.~\ref{sio}).  Multiple transitions exhibit similar double-peaked profiles ($\sim -5$ and $\sim 15$~\kms) and velocity extent (from $\sim -15$ to $\sim 30$~\kms). In addition, weaker features are observed in all profiles in the range $0 \rightarrow 10$~\kms. 

\subsubsection{$v=1, 2$ \te~lines: investigation of the possible existence of a high-velocity wind}
The \te~$v=1,2$ lines were our prime target within this survey. The spectral profiles are clearly dominated by two strong features, emitting in the range $-8 \rightarrow 0$~\kms (peak at $-5$~\kms) and $12 \rightarrow 23$~\kms (peak at 15~\kms) for $v=1$ and $-10 \rightarrow -4$~\kms (peak at $-7$~\kms) and $12 \rightarrow 23$~\kms (peak at 20~\kms) for $v=2$, respectively. 

\citet{Tan03} raise the possibility of high-velocity outflow during the later stages of massive star formation, when the protostar is expected to have contracted to the zero age main sequence while still in its main accretion stage. For example, the escape speed from the surface of such a 20~$M_\odot$ star is $\sim 1000$~\kms, and outflows launched from this vicinity are predicted to reach such speeds \citep{Shu00,Kon00}.
Because of the high temperatures and densities of environments at radii inside 10~AU, most common molecular tracers seen in star forming regions are unavailable. However, the maser emission from the \te~$v=1$ and $v=2$ transitions arises at temperatures of 1000-2000~K and densities of $10^{10 \pm 1}$~cm$^{-3}$, which makes SiO a plausible candidate as a tracer. Also, the intrinsically high brightness temperatures of maser emission enables detection from volumes that might be too small to generate detectable emission from thermal processes. With this in mind, we inspected the immediate vicinity of the vibrationally excited lines near the systemic velocity in the survey spectrum.

We found no obvious broad plateau (with a velocity extent of hundreds to thousands kilometer per seconds) or pattern of spectral features centered on the $v=1$ and $v=2$ emission, and  no compelling evidence for a high-velocity wind from Source~I. However, there are unidentified lines, at least some of which could in principle be high-velocity Doppler components of SiO. Prospective velocity offsets would be about --850, --170, 1000~\kms for $v=1$ and --120, 217, 440, 550~\kms for $v=2$, with intensities down by a factor of $\sim 3 \times 10^4$ from the respective emission peak. These would correspond to the outflow velocity projected onto the line of sight at different locations in the flow. Presumably, these would be where shocks have formed and where they are elongated along the line of sight (e.g., transverse to the direction of motion), so as to place the observer inside the maser beam.

We stress that suggested identifications, although appealing with regard to theoretical predictions, are highly speculative. In the absence of imaging data, the only possible assessment is via exclusion of other explanations. Among unidentified lines, 8 bracket the $v=2$ ($\nu=$42.742, 42.758, 42.790, 42.837~GHz) and the $v=1$ ($\nu=$42.979, 42.982, 43.146, 43.244~GHz) emission (see Fig.~\ref{spec}). Those at 42.790, 42.979, 42.982, 43.244~GHz do not correspond to any molecular transition in the reference catalogs. Near associations are possible for the remaining four, but ambiguity arises from conflicts among observed and anticipated line strengths.
At $\sim 42.742$~GHz, a transition is expected from H$_2$NCH$_2$CN, a rare species detected so far only towards Sagittarius B2(N) \citep{Bel08}. We discarded this identification because no other lines of similar expected emission characteristics are visible at higher frequencies in previous surveys. CH$_3$OCHO has a transition around 42.758~GHz, but the theoretical expected intensity is an order of magnitude lower than the line observed at 43.528~GHz, in disagreement with the two observed lines showing comparable strength.  At 42.837~GHz the isotopologue C$_4$C$^{13}$CH shows a line, but no other anticipated stronger lines are visible  within our band either from this or the main isotopologue, C$_6$H. Around 43.146~GHz a transition from C$_2$H$_4$O is expected, but the observed line appears to be blueshifted compared to the line observed at 43.3975~GHz and the  ratio of the two line intensities expected from laboratory measurements is inconsistent with the observations (0.007 vs 0.3).

 We explicitly note that, even if the unidentified lines were associated with  SiO, we would not be able to establish the location of the high-velocity  features using only single-dish observations. In principle, if the high-velocity flow is created near the protostar, it can still continue to large distances away from the protostar if there is no deceleration. \citet{Tay86} 
report 400~\kms line of sight velocities in the extended Orion-BN/KL outflow.
 Even in the hypothesis that the high-velocity SiO features would not arise in the proximity of Source~I but on large scales, we anticipate that they would still be spatially associated with Source~I, which is the only known origin of SiO emission in Orion BN/KL \citep{Beu05}.

\subsubsection{\tn~and \tr: first detection of the $v=0 \ J=1-0$ line}

The \tn~and \tr~$v=0\ J=1-0$ emission has been clearly detected toward Orion-BN/KL at 7~mm for the first time. Prior to this work, the \tr~$v=0\ J=1-0$ line was only marginally detected with the Nobeyama radio telescope by \citet{Cho98b}, and, to the best of our knowledge, no spectrum of the \tn~$v=0\ J=1-0$ line has previously been reported in the literature. In contrast, the \tn~and \tr~isotopologues are long known to emit the $v=0\ J=2-1$ line at 3~mm (first discovered by \citealt{Lov76} and \citealt{Wol80}, respectively).  

\begin{figure*}
\centering
\includegraphics[width=13cm]{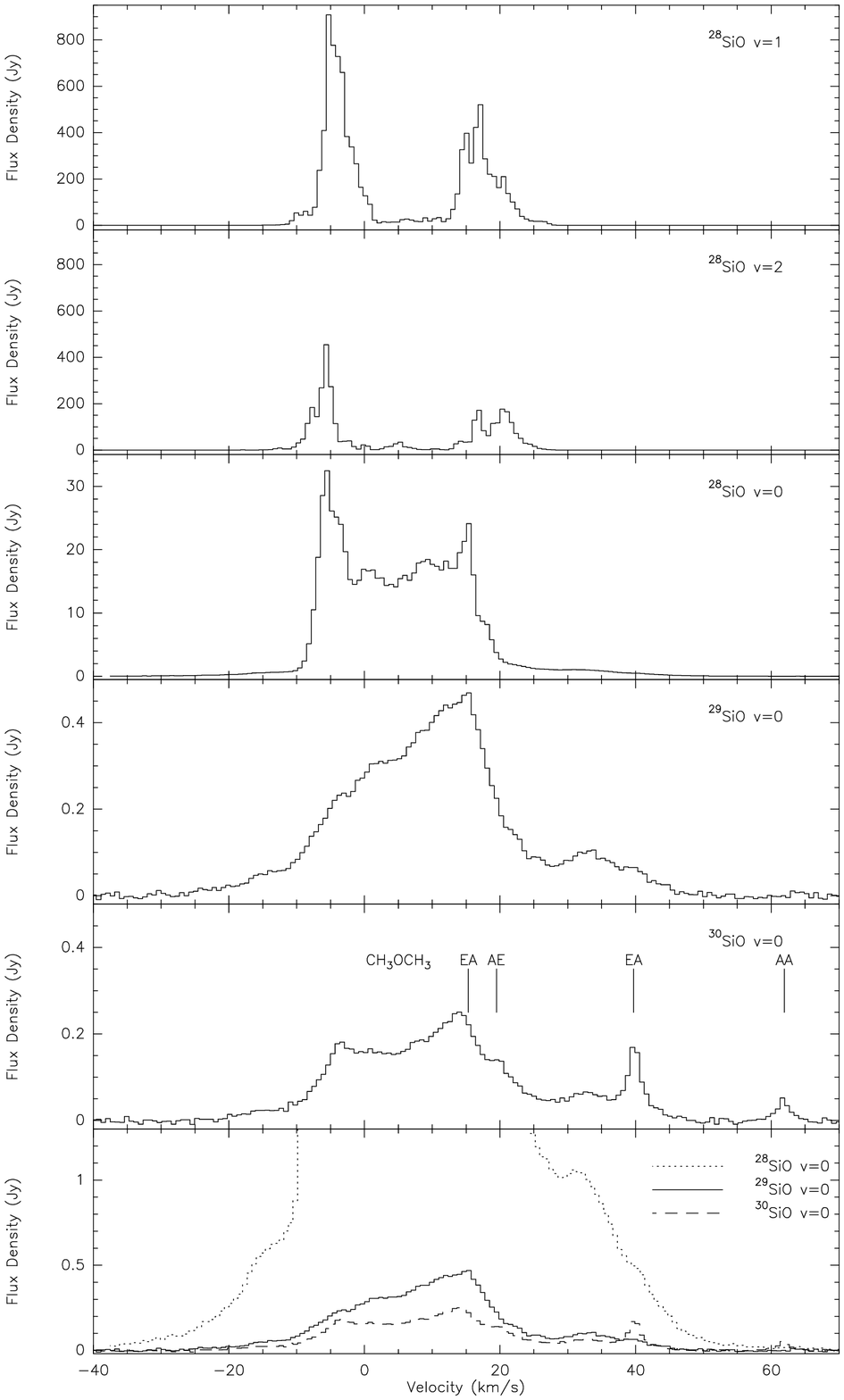}
\caption{Spectra of \te~$v=0,1,2 \ J=1-0$ and \tn~and \tr~$v=0 \ J=1-0$ from Orion-BN/KL observed with the GBT. 
The spectral resolution is 97.7~kHz (0.68~\kms) for all transitions. The vertical  lines in the \tr~profile   indicate the transitions of CH$_3$OCH$_3$ (see Table~\ref{lines}) corresponding to a velocity of 7~\kms with respect to the rest frequency of these transitions. In the bottom panel, $v=0$ line profiles of all isotopologues show similar redshifted (up to 50~\kms) and blueshifted (up to $-40$~\kms) wings of emission.
}
\label{sio}
\end{figure*}

 Both \tn~and \tr~have emission profiles similar to \te~$v=0$, with a flux density $\sim 200$ times  weaker. 
In particular, one unique characteristic of the $v=0$ emission (from all the three isotopologues) is a  wing of {\it redshifted} emission  extending to the range 30--50~\kms, not seen in the \te~$v=1,2$ profiles (Fig.~\ref{sio}, bottom panel). A weaker {\it blueshifted} wing is also present, from $-40$ to $-10$~\kms.
Previous single-dish observations at 3~mm showed broad thermal emission (from --25 to 40~\kms) for the \te~$v=0 \ J=2-1$ \citep{Olo81} and $J=3-2$ \citep{Cho99} lines. \tn~and \tr~$J=3-2$ lines show a similar, broad profile but the high-velocity wings fall below a $2\sigma$ level \citep{Cho99}. \citet{Cha95} imaged the $v=0\ J=1-0$ emission with the Very Large Array (VLA) and found low brightness redshifted (up to $\sim 42$~\kms) and blueshifted emission  (up to $\sim -22$~\kms). \citet{Gre04} proposed that the \te~$v=0$ traces a bipolar outflow along a northeast-southwest (NE-SW) direction at distances $\sim 200-1000$~AU from  Source~I. 
The similar breadths of wings for \tn~and \tr~relative to \te~suggests that emission from the former two molecules may also originate in the NE-SW bipolar outflow and  trace a moderately high-velocity component. On the other hand, \citet{Bau98} imaged the \te~$v=1$ and the \tn~$v=0$ $J=2-1$ emission with the Plateau de Bure Interferometer  and showed that both emissions arise within $\sim 100$~AU. This indicates the possibility that the \tn~$v=0$ (and \tr~as well) emission, similar to the \te~$v=1,2$ emission, might  trace instead the innermost region of the flows around Source~I. Imaging at high-angular resolution the $v=0 \ J=1-0$ emission of \te, \tn, and \tr~will be crucial in order to map the local gas dynamics around Source I.

 Looking at the \tr~profile in Fig.~\ref{sio}, one can note two narrow features overlapping the red  wing, unseen in the \te~and \tn~profiles.
 In fact, the rest frequencies of four transitions of CH$_3$OCH$_3$ (dimethylether) lie close to the rest frequency of \tr~(Fig.~\ref{sio}) and might be responsible for the redshifted line emission. 
Throughout the survey, dimethylether is found in 3 line groups (at $\sim 42.37,\ 42.45,\ 43.45$~GHz), consisting of four (resolved) torsional forms: $AA, EE, AE, EA$. In order to estimate the contribution of the CH$_3$OCH$_3$ lines of the first group to the redshifted portion of the \tr~profile, we performed a rotation diagram analysis on the lines of the second and third group, and then   used the estimated column density and rotational temperature to predict the line intensities of the first group. 
In Eq.~\ref{ntot}, the measured   integrated line intensities were weighted with the spin weight of the transition. Values for the intrinsic line strength, upper-state energy, spin weight, and  partition function ($91.62698T_{rot}^{3/2}$) were taken from  \citet{Gro98}.
 Assuming a source size of $10''$ to calculate the beam filling factor, the diagram yields $N_{tot} = 5.8 \times 10^{16}$~cm$^{-2}$ and $T_{rot} = 88$~K. 
 For the transitions of the first group, these values give the following line ratios: $AA:EE:AE:EA=3:8:2:1$.
 By comparing the observed and expected line ratios, a negligible contribution to the emission at 42.372~GHz is expected from CH$_3$OCH$_3$ (forms $AE,\ EA$). In contrast, the lines at 42.3656~GHz ($\sim 63$~\kms) and 42.3687~GHz  ($\sim 40$~\kms) are due to the $AA$ (100\%) and $EE$ (70\%) forms of dimethylether, with the plateau of emission under the  line at 40~\kms  probably due to \tr.
 Since  no other known transition falls in the range of frequencies from 42.366 to 42.370~GHz,  the redshifted wing of emission ($\sim 25-50$~\kms) is actually due to \tr, with the exception of the two lines at $\sim 40$ and $\sim 63$~\kms, consistent with the  \te~and \tn~profiles.

\subsubsection{Nature of the SiO emission at 7~mm}
 \te~$v=1,2$ transitions have been known to be inverted since their discovery \citep{Tha74,Buh74}. Based on brightness temperature arguments, \citet{Cha95} showed that \te~$v=0$ emission at 7~mm is part maser and part thermal (as is true for the 3~mm emission: \citealt{Wri95}).  
 \tn~and \tr~$v=0\ J=1-0$ emission could follow the same pattern.
 
In the case where a specific molecule emits part maser and part thermal emission, 
one would expect the thermal component to become dominant at higher frequencies {(e.g., \citealt{Loc92})}. In fact, in the sub-mm  regime ($\nu \sim 350$~GHz), interferometric imaging at high-angular ($\lesssim1''$) resolution  did not find any indication for \te~and \tr~($J=8-7$) maser emission \citep{Beu05}. On the other hand, \citet{Olo81} first proposed maser action for \tn~based on the observed variability of its spectral profile at 3~mm and \citet{Bau98} proved the maser nature of the \tn~$v=0 \ J=2-1$ emission based on interferometric  imaging.  These observational features at higher frequencies might be suggestive of maser emission for \tn~(and \tr) $v=0 \ J=1-0$ as well.

\clearpage
\clearpage
%
\subsection{CH$_3$COCH$_3$}
\label{ace}
Acetone is one of the most important molecules in organic chemistry. After 15 years of tentative interstellar identification, \citet{Sny02} confirmed its presence toward the hot molecular core Sagittarius B2(N-LMH).
The first detection of acetone in Orion BN/KL was reported by \citet{Fri05} at 3~mm.
Despite being structurally similar to CH$_3$OCH$_3$, a species commonly observed in SFRs, CH$_3$COCH$_3$ has not been reported in any other high-mass or low-mass SFR apart from Orion BN/KL and Sagittarius B2(N-LMH). Curiously as well, interferometric  maps show unequivocally that in Orion oxygen-bearing species such as dimthylether are observed towards the compact ridge, while  acetone is concentrated towards the hot core \citep{Fri05}.

In the present survey, we detected six transitions, confirming the identification of acetone at 7~mm. 
The lines not affected by blending were used to estimate  $N_{col}$ and $T_{rot}$ using the rotation diagram method (Eq.~\ref{ntot}). Similarly to dimethylether, the measured   integrated line intensities were weighted with the spin weight of the transition.
Following \citet{Fri05}, we adopted a rotational-vibrational partition function $261.7T_{rot}^{3/2} \times (1+e^{-115/T_{rot}}+e^{-180/T_{rot}})$ \citep{Gro02} and a source size of $5 \arcsec \times 3 \arcsec$ to calculate the beam filling factor. The values for the intrinsic line strength, upper state energy, and spin weight were taken from \citet{Gro02}.
Using the above quantities, the diagram yields $N_{tot} = 5.5 \times 10^{16}$~cm$^{-2}$ and $T_{rot} = 236$~K, in agreement  with the values estimated by \citet{Fri05} at 3~mm.

Both the high temperature and the central velocity of 5-7~\kms reported here are in agreement with  an origin of acetone in the hot core (typically T$\gtrsim150$~K). 

\clearpage
\begin{figure}
\centering
\includegraphics[width=0.5\textwidth]{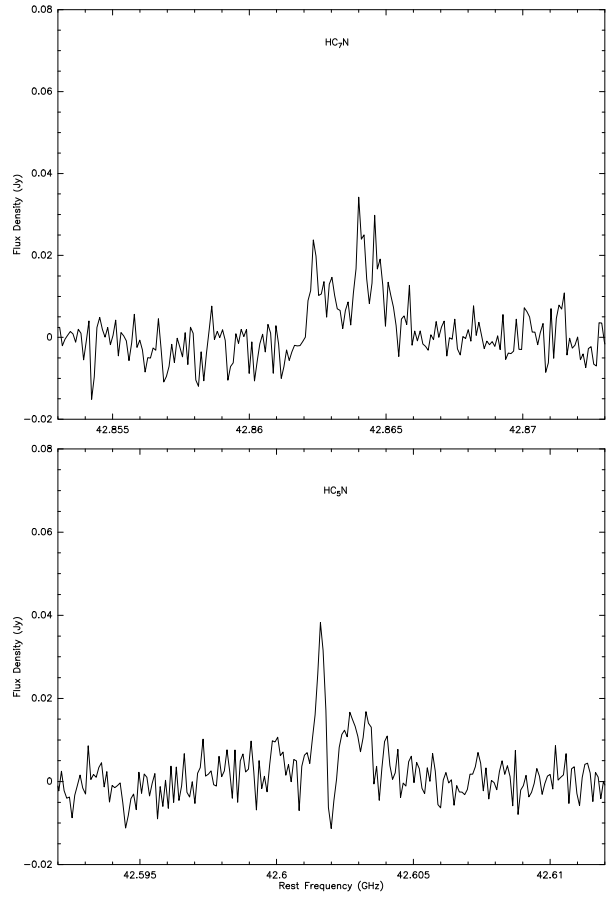}
\caption{Line profiles of the cyanopolyynes HC$_5$N ($16-15$) and HC$_7$N ($38_{38}-37_{37}$).}
\label{cyan}
\end{figure}
\clearpage

\subsection{Cyanopolyynes: HC$_5$N and HC$_7$N.}
\label{cyano}
Cyanopolyynes (HC$_{2n+1}$N, $n$ = 1, 2, 3, 4, 5, 6) are long c-chain species  believed to represent the early stages of chemical evolution in dense cores and so far found to be abundant (C/O$\sim$0.9) only towards the cold starless core TMC-1 \citep{Kai04}.
 Towards Orion BN/KL only a lighter species of cyanopolyynes, HC$_3$N, has been unambiguously identified in previous surveys and found to emit from the hot core and the extended ridge \citep{Tur91,Sch97}. \citet{Tur91} reported also detection of 10 and 8 transitions of HC$_5$N and HC$_7$N, respectively, but due to limited sensitivity only one line was detected above $3\sigma$ for each species (at 103.837~GHz and 73.316~GHz, respectively).
 
One transition of HC$_5$N ($J=16-15$) and of HC$_7$N ($J=38-37$), respectively, fall within our passband, and we detect them at the expected frequencies. In order to make the detection more robust, we checked for consistency with the results from \citet{Tur91}, by extrapolating the line intensities at 7~mm from the measurements at 3~mm and then comparing measured and predicted values.
One can in principle use Eq.~\ref{ntot} to predict the intensity of a specific transition, if the column density and rotational temperature of the molecule are known.  Adopting the values estimated by \citet{Tur91} from a rotation diagram (HC$_5$N: $T_{rot} \sim 31.2$~K, $N \sim 10^{13}$~cm$^{-2}$, HC$_7$N:   $T_{rot} \sim 40$~K, $N \sim 5 \times 10^{13}$~cm$^{-2}$), and taking $S\mu^2$, $E_u$, and $Q$ from the CDMS catalog, one derives $W\sim0.89$~K~\kms and  $W\sim0.98$~K~\kms for HC$_5$N and HC$_7$N, respectively. These values are consistent with those measured ($W\sim0.86$~K~\kms and  $W\sim1.1$~K~\kms), once the corrected antenna temperatures $T_a^*$ are corrected for the beam efficiency ($\eta_{MB}=0.6$) and the beam filling factor ($B=0.5$ in the assumption of extended source).

HC$_7$N shows a complex, triple-peaked profile (peaks 31, 44, and 39~mK) spanning 3~MHz (Fig.~\ref{cyan}), reflecting  the hyperfine structure of the  42.863~GHz transition. However, the predicted intensities for the side components are three orders of magnitude lower than the central component. 
 We speculate that this may be indicative of non-LTE conditions or limitations in applicability of laboratory measurements.
 It is improbable that the complex spectral structure of HC$_7$N is due to superposition of different velocity components in the region, because these show a maximum range of variation of $\sim 10$~\kms, whereas a 3~MHz span  corresponds to  21~\kms at 43~GHz. 
 We note that despite coarser resolution (1~MHz) and lower sensitivity, \citet{Tur91} also resolved the profile of the HC$_7$N line at 73.3~GHz detecting a FWHM of $\sim 5$~MHz or 21~\kms at 70~GHz, consistent with our finding.

The combined results of this survey and the work of \citet{Tur91} are strongly suggestive of the presence of heavy cyanopolyynes in Orion BN/KL. Long c-chain molecules, like cyanopolyynes, are believed to be abundant only in the early stage of chemical evolution of dark clouds and become depleted over time.
In the archetypical dark molecular cloud TMC-1 (T$_{kin}\sim 10$~K), complex species such as HC$_3$NH$^+$, C$_6$H, HC$_5$N, HC$_7$N, HC$_{9}$N, and HC$_{11}$N (the longest c-chain species discovered so far in the ISM)  are abundant \citep{Kai04,Bel97}. 
In this context, the detection of these species in Orion BN/KL, usually considered the archetype of  hot-core, may be surprising. However, based on the measured velocity and the low rotational temperature \citep{Tur91}, these species  are most probably associated with the extended ridge, which does consist of quiescent cold gas (T$_{kin}\lesssim 60$~K).

Finally, we explicitly note that there is no evidence in this survey of longer cyanopolyynes, in particular HC$_9$N and HC$_{11}$N (the latter detected so far only towards TMC-1).  Both species have three transitions in our passband but no emission is evident above $\sim 20$~mK ($3\sigma$). 
\citet{Rem06} studied spectral lines of different cyanopolyynes detected by \citet{Kai04} in TMC-1 and found  a linear inverse correlation  between species column density and number of carbon atoms in the chain. Based on this work, we anticipate an abundance three  and ten times lower
 for HC$_9$N and HC$_{11}$N, respectively, relative to the abundance for HC$_7$N. Assuming  a rotational temperature  similar to HC$_5$N/HC$_7$N (T$_{rot} \sim 35$~K - \citealt{Tur91}), we infer integrated intensities of the order of 0.1~K~\kms and 0.007~K~\kms for the most favorable transitions of HC$_9$N and HC$_{11}$N at 7~mm, close to or below our detection limit.
Laboratory measurements of rotational spectra at high frequencies ($\gtrsim 200$~GHz) complemented by sensitive surveys would be crucial for unambiguous identifications of these long c-chain species in the future.

\subsection{Other species of interest}
\label{others}
c-C$_2$H$_4$O$-$ Ethylene oxide is a high energy isomer of CH$_3$CHO (acetaldehyde).  \citet{Ike01} detected ethylene oxide towards Orion BN/KL for the first time, using the NRAO 12~m at Kitt Peak (1~mm) and the   NRO 45~m (3~mm). Only the most favorable line   ($8_{6,2}-8_{5,3}$ at 43397.5~MHz) among four in our passband is reported here, so no estimation of column density and rotational temperature can be made.
Acetaldehyde has been previously detected towards Orion BN/KL in several high-frequency surveys \citep{Sut85,Tur91,Ziu93}, however no lines of CH$_3$CHO fall in the band of this survey. 

HC3NH$^+ -$ We report here the possible detection of the $ \mathrm{HC_3NH^{+}}$ ion (protonated cyanoacetylene), which is thought to play an important role in the formation process to $ \mathrm{HC_3N}$ in molecular clouds  (see the network RATE06 at http://www.udfa.net/). It was detected for the first time toward TMC-1 using the Nobeyama 45~m radiotelescope at 7~mm (lines $J = 5-4$ and $J = 4-3$ - \citealt{Kaw94,Kai04}).  
Only the $J = 5-4$ transition falls in our passband. A transition from acetone falls close in frequency to the observed line, but the line center in that case would correspond to a blue-shifted velocity ($V_{LSR} \sim 0$~\kms), which is not observed in the other lines of acetone detected in this survey (having velocities in the range $\sim 5-7$~\kms).

\subsection{Undetected species in this survey}

Several species (C$_2$H$_3$CN, C$_2$H$_5$OH, CH$_3$OH, CH$_3$CN, HDS, NH$_3$, HDCO, H$_2$CS, HC$_3$N, C$_3$H$_2$, NH$_2$CN, HCS$^+$, HCO$^+$), some of which are known to be very abundant in Orion BN/KL and whose transitions dominate the spectrum  at mm and submm wavelengths \citep{Sch97,Sch01}, are not detected here although they have transitions in the passband. We now comment briefly on the implications of several of these non-detections.

 C$_2$H$_3$CN$-$ 
 24 transitions of C$_2$H$_3$CN are present in the  range of observed frequencies. The two  strongest transitions fall at 42.691~GHz and 43.345~GHz. Two lines are in fact observed at those frequencies, having a peak intensity of 23 and 80~mK, and a LSR velocity of $\sim 0$~\kms.  
However, C$_2$H$_3$CN is known to be abundant towards the hot core, with $V_{LSR}\sim 5$~\kms.
In addition, assuming a rotational temperature in the range 75-150~K, the expected ratio between the two line intensities would vary in the range 1-1.4, while 4 is actually observed. 
Extrapolations using the measurements by \citet{Sch97} at 330~GHz ($T_{rot}=96$~K, $N_{col}=8 \times 10^{14}$) predict peak and integrated intensities of $\sim 1$~mK and $\lesssim 20$~mK~\kms at 7~mm,  well below our detection limit. 
Based on the measured velocity and intensities, we suggest that the two lines are probably not associated with C$_2$H$_3$CN.
We attribute the line at 42.691~GHz to CH$_3$OCHO, which shows other two lines in the passband with consistent intensities.
 
HCS$^+ -$ 
A detection of the HCS$^+$ ion has been reported previously only by \citet{Sch97}. Based on the velocity and the width of the detected line, those authors proposed an origin in the extended ridge. 
In our passband, the $J=1-0$ line falls  at 42674.19~MHz, where it is blended with the $11_{1,10}-11_{0,11}$ line from  C$_2$H$_5$CN. 
Using a rotation diagram, we infer $T_{rot}=87$~K and $N_{col}=1.2 \times 10^{17}$~cm$^{-2}$ for  C$_2$H$_5$CN. Using these values, we predict an integrated  intensity of 1.4~K~\kms for the $11_{1,10}-11_{0,11}$ line , in agreement with the measured value 1.2~K~\kms. Hence, we attribute the detected emission at 42674.2~MHz to  C$_2$H$_5$CN. 

CH$_3$OH $-$
Four transitions of methanol are present in the passband with $E_u \sim 500-1300$~K. CH$_3$OH dominates the spectrum at higher frequencies and the observed transitions have upper state energy levels of $\sim 50$ to $\gtrsim 700$~K over 70-900~GHz. Adopting $T_{rot}=190$~K and $N_{col}=7 \times 10^{16}$ \citep{Sch97}, for the transitions at 7~mm one would infer integrated intensities too weak to be detected ($\lesssim 3$~mK~\kms).

C$_2$H$_5$OH $-$
Thirteen transitions of ethanol are present in the  range of observed frequencies, but none of them was detected. Indications of the presence of ethanol in Orion-KL has been previously reported \citep{Ziu93,Sut95,Sch97}, but in all of these surveys the ethanol lines were so weak that they could not be identified conclusively. The existence of ethanol has been proved by \citet{Ohi95}, via detection of four transitions lying 13-31~K above the ground state with the NRO 45-m telescope in the frequency range 81.7-104.8~GHz. Almost all transitions present in the frequency range of our survey have $E_u \gtrsim 100$~K.  Based on the intensities observed by \citet{Ohi95} at 3~mm, the most favorable transition ($J=1-0$) is expected to be below our detection limit ($\lesssim10$~mK).

CH$_3$CN $-$
CH$_3$CN has five transitions with $E_u > 800$~K. Although high energy levels (up to 1000~K) have been detected in the $\sim 900$~GHz band \citep{Com05}, extrapolations based on those data predict  intensities too weak ($\lesssim 0.1$~mK) to be detected here.

Others $-$ The remaining species  have only one transition with $E_u \gtrsim 1000-1500$~K (H$_2$CS has $E_u = 550$~K and an expected intensity $\lesssim 0.1$~mK), resulting in no favorable transition at observed frequencies. 

\subsection{Radio recombination lines}
\label{rec}
Three radio recombination lines (RRLs) are detected in the survey: H76$\gamma$ ($\nu=42.414$~GHz),  H53$\alpha$ ($\nu=42.952$~GHz), and  He53$\alpha$ ($\nu=42.969$~GHz).
Among all the lines observed in the survey, the RRLs show the most blue-shifted emission with velocities ranging from  $-5$ to $-10$~\kms. Previous surveys show that line-center velocity increases  with transition frequency: observations of several lines (from H46$\beta$, H44$\alpha$ to H39$\alpha$, H48$\beta$) in the frequency range 71-122~GHz provided a  mean velocity $v_{LSR}=-3.4$~\kms \citep{Tur91}, H43$\beta$ at $\nu=154.6$~GHz had $v_{LSR}$=2.4~\kms, and finally H30$\alpha$ at $\nu=230$~GHz had $v_{LSR}$=4~\kms  \citep{Sut85}. This effect is observed in \HII \ regions with   density gradients and may be explained by the correlation of density with observing frequency: higher frequency lines are associated with higher density gas. Following \citet{Ket95}, the peak intensity of RRLs is proportional to  density squared and inversely proportional to  line width. At low frequency the line width is dominated by pressure broadening, which is proportional to  gas density. As a consequence,  peak intensity of low frequency lines are proportional to  density. For high frequency lines (with presumably negligible pressure broadening),  peak intensity is proportional to  density squared. So, if either multiple sources or a density gradient within a single source is present, high-frequency RRLs will trace predominantly high density gas, while low-frequency RRLs will trace both high- and low-density gas. 
Hence, the observed difference in velocity between high- and low-frequency RRLs in Orion BN/KL probably indicates a difference in velocity between  high and low-density gas. The systematic increase of the line center velocity with frequency can be explained with the presence of both a density gradient and an outflow/inflow in the \HII \ region.

\section{Summary}
In this line survey of the Orion BN/KL region we find 67 spectral features, attributed to 13 species.
The spectrum is dominated by SiO, SO$_2$, C$_2$H$_5$CN, CH$_3$OCH$_3$, and CH$_3$COCH$_3$. 
We confirm the detection of acetone towards the hot core at 7~mm, and we find further indications of the association of long c-chain molecules with the extended ridge: HC$_5$N, HC$_7$N, and possibly the cation HC$_3$NH$^+$. The detections of these c-chain molecules are based only on single transitions, so follow-up observations are necessary to obtain confirmation.

Five transitions from SiO isotopologues have been detected. The profiles of the strong \te~$v=1,2$ maser transitions extend in the velocity range from $\sim -20$ to $+30$~\kms.
 Some of the unidentified lines in principle could be  high-velocity (100-1000~\kms) Doppler components of SiO, but without imaging data the location of the presumed high-velocity features cannot be established. Future observations with the NRAO EVLA interferometer, providing broad instantaneous bandwidth, tuning flexibility, and high sensitivity, will be crucial 
   to establish the origins of the candidate high-velocity emission. 

 We report for the first time spectral profiles of the line $v=0\ J=1-0$ from \tn~and \tr, which overlap well with the velocity range of  the $v=0$ \te~ emission.  In particular, the profiles of the $v=0$ emission from all the  \te, \tn~and \tr~isotopologues show a redshifted wing of emission extending up to  50~\kms (and a less prominent blueshifted wing up to --40~\kms). 
  We speculate that the wing emission  may trace a moderately high-velocity wind component from Source~I, but with only single-dish spectra no firm conclusion can be drawn about the origin of the emission. 
 Imaging at high-angular resolution the $v=0$ emission from all the SiO isotopologues will be uniquely important in the mapping  of the local gas dynamics. That would also be  essential to establish without ambiguity the nature (thermal vs maser) of the $v=0$ emission from \tn~and \tr. 
 Indeed, with the aim of establishing the origin and the nature of the \tn~and \tr~$v=0\ J=1-0$ emission, we conducted follow-up observations of Orion-BN/KL in the three isotopologues with the NRAO VLA interferometer; the results will be discussed in a  forthcoming paper.
 The presence of two vibrationally excited transitions and three isotopologues from SiO, makes Orion-BN/KL the only laboratory for studying excitation mechanisms of isotopic SiO masers in SFRs (and possibly in late-type stellar envelopes).

\acknowledgments
We wish to thank R. Maddalena, D. Balser, and T. Minter for helpful discussions on GBT calibration parameters. We thank M. Stennes and T. Minter  for analyses that explained the spectral artifacts  reported here.
We are also very grateful to the AMP spectroscopy group at the CfA, in particular S. Brunken and M. McCarthy, for very helpful discussions about some identifications reported in the present survey. The data presented here were obtained under the GBT program 07A-102.
This material is based upon work supported by the National Science Foundation under Grant No. NSF AST 0507478.
%
%


\begin{thebibliography}{}

\bibitem[Allen 
\& Burton(1993)]{All93} Allen, D.~A., \& Burton, M.~G.\ 1993, \nat, 363, 54
\bibitem[Bally 
\& Zinnecker(2005)]{Bal05} Bally, J., \& Zinnecker, H.\ 2005, \aj, 129, 2281  
\bibitem[Baudry et 
al.(1998)]{Bau98} Baudry, A., Herpin, F., \& Lucas, R.\ 1998, \aap, 335, 654 
\bibitem[Bell et al.(1997)]{Bel97} Bell, M.~B., Feldman, 
P.~A., Travers, M.~J., McCarthy, M.~C., Gottlieb, C.~A., 
\& Thaddeus, P.\ 1997, \apjl, 483, L61 
\bibitem[Belloche et al.(2008)]{Bel08} Belloche, A., Menten, K.~M., Comito, C., M{\"u}ller, H.~S.~P., Schilke, P., Ott, J., Thorwirth, S., \& Hieret, C.\ 2008, \aap, 482, 179 
\bibitem[Beuther et al.(2005)]{Beu05} Beuther, H., et al.\ 
2005, \apj, 632, 355
\bibitem[Beuther et al.(2006)]{Beu06} Beuther, H., et al.\ 
2006, \apj, 636, 323 
\bibitem[Blake et al.(1986)]{Bla86} Blake, G.~A., Masson, 
C.~R., Phillips, T.~G., \& Sutton, E.~C.\ 1986, \apjs, 60, 357
\bibitem[Blake et al.(1996)]{Bla96} Blake, G.~A., Mundy, 
L.~G., Carlstrom, J.~E., Padin, S., Scott, S.~L., Scoville, N.~Z., 
\& Woody, D.~P.\ 1996, \apjl, 472, L49
\bibitem[Bonnell et al.(1998)]{Bon98} Bonnell, I.~A., Bate, 
M.~R., \& Zinnecker, H.\ 1998, \mnras, 298, 93 
\bibitem[Bonnell 
\& Bate(2006)]{Bon06} Bonnell, I.~A., \& Bate, M.~R.\ 2006, \mnras, 370, 488 
\bibitem[Buhl et al.(1974)]{Buh74} Buhl, D., Snyder, L.~E., 
Lovas, F.~J., \& Johnson, D.~R.\ 1974, \apjl, 192, L97 

\bibitem[Chandler 
\& de Pree(1995)]{Cha95} Chandler, C.~J., \& de Pree, C.~G.\ 1995, \apjl, 455, L67 \bibitem[Charnley(2004)]{Cha04} Charnley, S.~B.\ 2004, 
Advances in Space Research, 33, 23
\bibitem[Cho et al.(1998)]{Cho98} Cho, S.-H., Chung, H.-S., 
Kim, H.-R., Oh, B.-Y., Lee, C.-H., \& Han, S.-T.\ 1998, \apjs, 115, 277 
\bibitem[Cho 
\& Ukita(1998)]{Cho98b} Cho, S.-H., \& Ukita, N.\ 1998, \aj, 116, 2495 
\bibitem[Cho et al.(1999)]{Cho99} Cho, S.-H., Chung, H.-S., 
Kim, H.-R., Kim, H.-G., \& Roh, D.-G.\ 1999, \aj, 117, 1485 
\bibitem[Comito et al.(2005)]{Com05} Comito, C., Schilke, P., 
Phillips, T.~G., Lis, D.~C., Motte, F., 
\& Mehringer, D.\ 2005, \apjs, 156, 127 
\bibitem[Friedel et al.(2005)]{Fri05} Friedel, D.~N., et al.\ 
2005, \apj, 630, 623 
\bibitem[Genzel et al.(1981)]{Gen81} Genzel, R., Reid, M.~J., 
Moran, J.~M., \& Downes, D.\ 1981, \apj, 244, 884 
\bibitem[Genzel 
\& Stutzki(1989)]{Gen89} Genzel, R., \& Stutzki, J.\ 1989, \araa, 27, 41 
\bibitem[Greaves 
\& White(1991)]{Gre91} Greaves, J.~S., \& White, G.~J.\ 1991, \aaps, 91, 237 
\bibitem[Greenhill et al.(2004)]{Gre04} Greenhill, L.~J., 
Reid, M.~J., Chandler, C.~J., Diamond, P.~J., 
\& Elitzur, M.\ 2004, Star Formation at High Angular Resolution, 221, 155 
\bibitem[Groner et al.(1998)]{Gro98} Groner, P., Albert, S., 
Herbst, E., \& De Lucia, F.~C.\ 1998, \apj, 500, 1059 
\bibitem[Groner et al.(2002)]{Gro02} Groner, P., Albert, S., 
Herbst, E., De Lucia, F.~C., Lovas, F.~J., Drouin, B.~J., 
\& Pearson, J.~C.\ 2002, \apjs, 142, 145 
\bibitem[Ikeda et al.(2001)]{Ike01} Ikeda, M., Ohishi, M., 
Nummelin, A., Dickens, J.~E., Bergman, P., Hjalmarson, {\AA}., 
\& Irvine, W.~M.\ 2001, \apj, 560, 792  
\bibitem[Jewell et al.(1989)]{Jew89} Jewell, P.~R., Hollis, 
J.~M., Lovas, F.~J., \& Snyder, L.~E.\ 1989, \apjs, 70, 833 
\bibitem[Johansson et 
al.(1984)]{Joh84} Johansson, L.~E.~B., et al.\ 1984, \aap, 130, 227 

\bibitem[Kaifu et al.(2004)]{Kai04} Kaifu, N., et al.\ 2004, 
\pasj, 56, 69 

\bibitem[Kawaguchi et al.(1994)]{Kaw94} Kawaguchi, K., Kasai, 
Y., Ishikawa, S.-I., Ohishi, M., Kaifu, N., 
\& Amano, T.\ 1994, \apjl, 420, L95
\bibitem[Keto et al.(1995)]{Ket95} Keto, E.~R., Welch, W.~J., 
Reid, M.~J., \& Ho, P.~T.~P.\ 1995, \apj, 444, 765 
 \bibitem[Konigl 
\& Pudritz(2000)]{Kon00} Konigl, A., \& Pudritz, R.~E.\ 2000, Protostars and Planets IV, 759 


\bibitem[Lee et al.(2001)]{Lee01} Lee, C.~W., Cho, S.-H., 
\& Lee, S.-M.\ 2001, \apj, 551, 333 
\bibitem[Lee 
\& Cho(2002)]{Lee02} Lee, C.~W., \& Cho, S.-H.\ 2002, Journal of Korean Astronomical Society, 35, 187 
\bibitem[Lockett \& Elitzur(1992)]{Loc92} Lockett, P., \& Elitzur, M.\ 1992, \apj, 399, 704 
\bibitem[Lovas et al.(1976)]{Lov76} Lovas, F.~J., Johnson, 
D.~R., Buhl, D., \& Snyder, L.~E.\ 1976, \apj, 209, 770 
\bibitem[Lovas(2004)]{Lov04} Lovas, F.~J.\ 2004, Journal of 
Physical and Chemical Reference Data, 33, 177 

\bibitem[McKee 
\& Tan(2003)]{Mck03} McKee, C.~F., \& Tan, J.~C.\ 2003, \apj, 585, 850 
\bibitem[Menten et al.(2007)]{Men07} Menten, K.~M., Reid, M.~J., Forbrich, J., \& Brunthaler, A.\ 2007, \aap, 474, 515
\bibitem[Minter(2007)]{Min07a} Minter, A. H. 2007, GBT Memo no. 247
\bibitem[Minter \& Stennes(2007)]{Min07b} Minter, A. H. \& Stennes, M. J. 2007, GBT Memo no. 257

\bibitem[M{\"u}ller et 
al.(2001)]{Mul01} M{\"u}ller, H.~S.~P., Thorwirth, S., Roth, D.~A., \& Winnewisser, G.\ 2001, \aap, 370, L49

\bibitem[Nikolic et 
al.(2007)]{Nik07} Nikolic, B., Prestage, R.~M., Balser, D.~S., Chandler, C.~J., \& Hills, R.~E.\ 2007, \aap, 465, 685
\bibitem[Ohishi et al.(1995)]{Ohi95} Ohishi, M., Ishikawa, 
S.-I., Yamamoto, S., Saito, S., \& Amano, T.\ 1995, \apjl, 446, L43 
\bibitem[Olofsson et 
al.(1981)]{Olo81} Olofsson, H., Hjalmarson, A., \& Rydbeck, O.~E.~H.\ 1981, \aap, 100, L30
 \bibitem[Olofsson et 
al.(2007)]{Olo07} Olofsson, A.~O.~H., et al.\ 2007, \aap, 476, 791 
\bibitem[Pickett et al.(1998)]{Pic98} Pickett, H.~M., 
Poynter, I.~R.~L., Cohen, E.~A., Delitsky, M.~L., Pearson, J.~C., 
\& Muller, H.~S.~P.\ 1998, Journal of Quantitative Spectroscopy and Radiative Transfer, 60, 883 

\bibitem[Remijan et al.(2006)]{Rem06} Remijan, A.~J., Hollis, 
J.~M., Snyder, L.~E., Jewell, P.~R., \& Lovas, F.~J.\ 2006, \apjl, 643, L37 

\bibitem[Schilke et al.(1997)]{Sch97} Schilke, P., Groesbeck, 
T.~D., Blake, G.~A., \& Phillips, T.~G.\ 1997, \apjs, 108, 301 
\bibitem[Schilke et al.(2001)]{Sch01} Schilke, P., Benford, 
D.~J., Hunter, T.~R., Lis, D.~C., \& Phillips, T.~G.\ 2001, \apjs, 132, 281
\bibitem[Shu et al.(2000)]{Shu00} Shu, F.~H., Najita, J.~R., 
Shang, H., \& Li, Z.-Y.\ 2000, Protostars and Planets IV, 789 
\bibitem[Snyder et al.(2002)]{Sny02} Snyder, L.~E., Lovas, 
F.~J., Mehringer, D.~M., Miao, N.~Y., Kuan, Y.-J., Hollis, J.~M., 
\& Jewell, P.~R.\ 2002, \apj, 578, 245 
\bibitem[Snyder et al.(2005)]{Sny05} Snyder, L.~E., et al.\ 
2005, \apj, 619, 914 
\bibitem[Sutton et al.(1985)]{Sut85} Sutton, E.~C., Blake, 
G.~A., Masson, C.~R., \& Phillips, T.~G.\ 1985, \apjs, 58, 341
\bibitem[Sutton et al.(1995)]{Sut95} Sutton, E.~C., Peng, R., 
Danchi, W.~C., Jaminet, P.~A., Sandell, G., 
\& Russell, A.~P.~G.\ 1995, \apjs, 97, 455 
\bibitem[Tan 
\& McKee(2003)]{Tan03} Tan, J.~C., \& McKee, C.~F.\ 2003, ArXiv Astrophysics e-prints, arXiv:astro-ph/0309139 
\bibitem[Tan(2004)]{Tan04} Tan, J.~C.\ 2004, \apjl, 607, L47
\bibitem[Taylor et al.(1986)]{Tay86} Taylor, K., Dyson, 
J.~E., Axon, D.~J., \& Hughes, S.\ 1986, \mnras, 221, 155 
\bibitem[Thaddeus et al.(1974)]{Tha74} Thaddeus, P., Mather, 
J., Davis, J.~H., \& Blair, G.~N.\ 1974, \apjl, 192, L33 
\bibitem[Turner(1989)]{Tur89} Turner, B.~E.\ 1989, \apjs, 70, 
539  
\bibitem[Turner(1991)]{Tur91} Turner, B.~E.\ 1991, \apjs, 76, 
617 
\bibitem[White et 
al.(1986)]{Whi86} White, G.~J., Griffin, M.~J., Rainey, R., Monteiro, T.~S., \& Richardson, K.~J.\ 1986, \aap, 162, 253 
\bibitem[White et 
al.(2003)]{Whi03} White, G.~J., Araki, M., Greaves, J.~S., Ohishi, M., \& Higginbottom, N.~S.\ 2003, \aap, 407, 589 
\bibitem[Wolff(1980)]{Wol80} Wolff, R.~S.\ 1980, \apj, 242, 
1005 
\bibitem[Wright et al.(1995)]{Wri95} Wright, M.~C.~H., Plambeck, R.~L., Mundy, L.~G., \& Looney, L.~W.\ 1995, \apjl, 455, L185 
\bibitem[Ziurys \& McGonagle(1993)]{Ziu93} Ziurys, L.~M., \& McGonagle, D.\ 1993, \apjs, 89, 155 


\end{thebibliography}
\end{document}